\documentclass[12pt]{article}
\usepackage{graphicx}

\input epsf
\def\beq{\begin{eqnarray}}
\def\eeq{\end{eqnarray}}

\def\k{{\bf k}}

\def\lsim{\mathrel{\rlap{\lower3pt\hbox{\hskip0pt$\sim$}}
     \raise1pt\hbox{$<$}}}         
\def\gsim{\mathrel{\rlap{\lower4pt\hbox{\hskip1pt$\sim$}}
     \raise1pt\hbox{$>$}}}         

\usepackage{amsmath}
\usepackage{amsfonts}
\usepackage{verbatim}

\begin{document}


\thispagestyle{empty}

\begin{flushright}
{NYU-TH-09/06/15}
\end{flushright}
\vskip 0.9cm

\centerline{\Large \bf Strongly Coupled Condensate of High Density Matter}
\vskip 0.2cm
\centerline{\Large \bf }                    

\vskip 0.7cm
\centerline{\large Gregory Gabadadze}
\vskip 0.3cm
\centerline{\em Center for Cosmology and Particle Physics,  
Department of Physics,}
\centerline{\em New York University, New York, 
NY  10003, USA}

\vskip 1.9cm

\begin{abstract}

Arguments are summarized, that neutral matter made of  helium, carbon, etc., 
should form a quantum liquid at the above-atomic but below-nuclear densities
for which the charged spin-0 nuclei can condense.
The resulting substance has distinctive features, such as a mass gap in the 
bosonic sector and a gap-less spectrum of quasifermions, which 
determine its thermodynamic properties. I discuss an effective field theory description
of this substance, and as an  example,  consider its application to calculation of a 
static potential between  heavy charged impurities. The potential 
exhibits a long-range oscillatory behavior in which both the fermionic and  bosonic 
low-energy degree of freedom contribute. Observational consequences of 
the condensate for cooling of helium-core white dwarf stars are briefly discussed.

\end{abstract}

\vspace{2cm}

\begin{center}
{\it Based on a talk given at the international workshop \\
``Crossing the boundaries: Gauge dynamics at strong coupling'' \\
honoring the 60th birthday of M.A. Shifman\\
 Minneapolis, May 14-17, 2009}

\end{center}


\newpage

\subsection*{Foreword}

Like many in the audience, I first met Misha  on  the pages of journal 
publications, before meeting him in person.  While working on an undergraduate  
thesis at Moscow University, I came across  
Misha's review paper ``Anomalies and Low-Energy Theorems of Quantum 
Chromodynamics'' \cite {1}. Impressions of that work were very distinct -- 
a clear  exposition  of subtle field theory aspects of 
the quantum anomalies,  culminating in creative applications to low-energy 
hadron phenomenology. The work stood out by its originality, 
depth, inspiration and balance of the formalism and applications -- 
the remarkable signatures of Misha's enormous contribution to 
theoretical physics  at the forefront of both field 
theory and  particle phenomenology.
 
I met Misha in person in Minneapolis in 1998. The discussion with him was very 
inspiring.  Soon,  in Aspen,  we started to work on a project. A bit later I ceased  the  
opportunity to get exposed to two years of a unique  FTPI experience. We continued 
to work on and off on various projects since then. I value those works very highly, and  
feel privileged, as I'm sure many of you do too, for having such a collaborator. 

\vspace{0.1cm}

Happy 60th Birthday Misha!

\subsection*{Description of charged condensate}

Consider a neutral system of a large number of nuclei each having  
charge $Z$, and neutralizing electrons. If average inter-particle  separations 
in this system are  much smaller than  the atomic scale, $ \sim 10^{-8}~cm$,   
while being much larger than the nuclear scale,  $ \sim 10^{-13}~cm$, neither the 
atomic  nor nuclear effects  will play  any significant role. 
Moreover, the nuclei can also be treated as point-like particles. 

In what follows we focus on spin-0 nuclei with $Z\leq 8$ (helium, carbon, oxygen), 
and consider the electron number-density in the interval 
$J_0\simeq (0.1-5~MeV)^3$. Then the electron Fermi energy 
will exceed the electron-electron and electron-nucleus 
Coulomb interaction energy. Moreover, at temperatures below $\sim 10^7~K$, which are 
of interest here, the system of electron represents  
a degenerate Fermi gas.

Since the nuclei (we also  call them ions below) are heavier, 
temperature at which they'll start to exhibit quantum properties will be lower. 
Let us define  the ``critical''  temperature  $T_c$, at which the de 
Broglie wavelengths of the ions begin  to overlap
\begin{equation}
T_c\simeq\frac{4\pi^2}{3m_Hd^2}\,,~~~~~d \equiv 
\left ( 3 Z \over 4 \pi   J_0 \right )^{1/3}\,,
\label{Tc}
\end{equation}
where, $m_H$ denotes the mass of the ion (the subscript $''H''$ 
stands for heavy), and $d$ denotes the average separation  
between the ions\footnote{The de Broglie wavelength above is defined 
as $\lambda_{dB}=2\pi/|\k|$,
where $\k^2/2m_H = 3k_B T/2$. We define $T_c$ as the temperature at which 
 $\lambda_{dB}\simeq d$. Note that this differs by a numerical 
factor of $\sqrt{2\pi/3}$ from the standard definition of the 
{\it thermal} de Broglie  wavelength,  $\Lambda \equiv \sqrt{ 2\pi /mk_BT}$,  
that appears  in the partition function of an ideal gas 
of number-density $n$ in the dimensionless combination $\Lambda^3 n$.}.

Somewhat below $T_c$ quantum-mechanical uncertainties in the ion positions become 
greater than  an average inter-ion separation. Hence the latter concept 
looses its meaning as a microscopic characteristic  of the system;   
the ions enter a quantum-mechanical regime of indistinguishability.
Then,  the many-body wavefunction of the spin-0 ions should be 
symmetrized, and this would  unavoidably lead to probabilistic ``attraction''  
of the bosons to condense, i.e., to occupy one and the same quantum state.
We refer the system of condensed nuclei and electrons as charged condensate.

In the  condensate the scalars occupy a quantum state with zero  
momentum.  Moreover, small fluctuations of  the bosonic  
sector happen  to have a mass gap,  $m_\gamma = (Z e^2 J_0/m_H)^{1/2}$,  
which exceeds  $T_c$ by more than  an order of magnitude. 
Therefore, once bosons are in the charged condensate,  
their phonons cannot be thermally excited. 
However,  the gap-less fermionic  degrees of freedom are  
thermally excited, and carry the most of the entropy 
of the entire system  \cite {GGRR1}-\cite{GGDP}.

For further discussions it is useful to rewrite the expression for 
$T_c$ in terms of the mass density $\rho\equiv m_HJ_0$ measured in $g/cm^3$:
\beq
T_c =  \rho^{2/3} \, \left ( {3.5 \cdot 10^2\over Z^{5/3}}\right )~K\,,
\label{Tcrho}
\eeq
where the baryon number of an ion was assumed to equal twice 
the number of protons, $A=2Z$ (true for helium, carbon, oxygen...). 
Thus, for $\rho =10^6~g/cm^3$  and  helium-4 nuclei we get
$T_c \simeq 10^6~K$, while for the carbon nuclei with the same mass density 
$T_c \simeq 2\cdot 10^5~K$.

Temperature at which the condensation phase transition takes place, $T_{condens}$,  
need not coincide with $T_{c}$. Moreover, we would expect
$T_{condens}\lsim T_c$. Calculation of $T_{condens}$ 
from the fundamental principles of this theory is hard. However, we can obtain 
an interval in which  $T_{condens}$ should fit. 
For this we  introduce the following parametrization:
\beq
T_{condens}= \zeta\, T_c\,,
\label{Tcond}
\eeq
where $\zeta$ is an unknown  dimensionless parameter that should depend  on 
density more mildly than  $T_c$ does.  Numerically, however, this 
parameter should vary in the interval 
$0.1 \ll  \zeta \lsim 1$:  The point $\zeta=0.1$ would  corresponds to the 
temperature of the 
Bose-Einstein (BE) condensation of a free gas  for which, 
$T^{BE}_{condens}\simeq 1.3/m_Hd^2$,  is known from the fundamental principles. 
The condensation temperature in our system should be higher than $T^{BE}_{condens}$
since the repulsion makes easier for the condensation to take place \cite {Huang}. 
In our case, repulsive interactions between the bosons are {\it strong} --
the Coulomb energy is at  least an order of magnitude greater that any other energy 
scale in the system. Hence, we should expect $\zeta \gg 0.1$.  On the other hand,  
given the definition of $T_c$, the parameter $\zeta$  cannot be greater than unity.
In what follows we will retain $\zeta $ in our expressions, 
but  use $\zeta \simeq 1$ when it comes to numerical estimates.

\vspace{0.1cm}

The condensation will take place after gradual cooling,  
only if $T_{condens}$ is greater than the temperature at which the 
substance could crystallize.  A classical plasma crystallizes 
when the Coulomb energy becomes about $\sim 180$  times greater than  
the average thermal energy per particle \cite {vanHorn,Ichi,DeWitt}. 
This gives the following crystallization temperature\footnote{The presented 
formula for the crystallization temperature  
is entirely {\it classical}. The temperature scale that  determines the 
classical versus quantum nature of the 
crystallization transition is the Debye temperature
$\theta_D\simeq4\cdot10^3\rho^{1/2}~K$.
Often, $\theta_D$ may  significantly exceed  
$T_{\text{cryst}}$ \cite{Ashcroft}. In such cases, 
quantum zero-point oscillations should be taken into account. 
This seems to delay the formation of quantum crystal, lowering 
$T_{\text{cryst}}$ from its classical value at most by about 
$\sim 10 \% $ \cite{Chabrier}.  Since this is a small change, we will ignore it 
in our estimates.}
\begin{equation}
T_{\text{cryst}}\simeq \rho^{1/3} \left 
(0.8\cdot 10^3 Z^{5/3} \right )~K\,.
\label{Tcryst}
\end{equation}
Note that the density dependence of $T_c$ is different from that 
of $T_{\text{cryst}}$ --  for higher densities $T_c$ grows faster, making 
condensation more and more favorable! One can define the ``equality'' density 
for which  $T_{condens}=T_{\text{cryst}}$:
\beq
\rho_{\rm eq}= \left ( {2.3\over \zeta}\right )^3 Z^{10}\,g/cm^3\,.
\label{equality}
\eeq
For helium, $Z=2$,  and $\rho_{\rm eq}\simeq 10^4~g/cm^3$, 
while for carbon, $Z=6$, and  $\rho_{\rm eq}\simeq  10^9~g/cm^3$ 
(as mentioned above, we use  $\zeta\simeq 1$). These results  
are very sensitive to the value of  $\zeta$; for instance, $\rho_{\rm eq}$  
could be an order of magnitude higher if $\zeta \simeq 0.5$. Irrespective of this 
uncertainty, however, the obtained densities are in the right ballpark of 
average densities  present in helium-core white dwarfs $\sim  10^6~g/cm$, 
(for carbon dwarfs,  they're closer to those expected in high 
density regions only \cite {GGDP}.)


Is the charged condensate a ground state  of the system at hand? 
For the higher values of the density interval considered, the crystal would 
not exist due to strong zero-point oscillations. At lower densities, the 
crystalline  state has lower free energy (at least near 
zero temperature) due to more favorable Coulomb binding. 
Hence, the condensate can only be a metastable state. The  
question arises whether after condensation at  $\sim T_{condens}$ 
the system could transition at lower temperatures 
$\sim T_{\rm cryst}$ to   the crystal state, 
as soon as the latter becomes available.

In the condensate, the boson positions are entirely uncertain  
while their momenta equal to zero. In order for such a system  
to crystallize later on, each of the bosons should acquire 
energy of the zero-point oscillations of  
crystal ions. As long as this energy, $\sim (Z e^2 J_0/m_H)^{1/2}$,  
is much greater than  $T_{\rm cryst}$, no thermal fluctuations can excite the 
condensed bosons to transition to the crystalline state. The latter 
condition is well-satisfied for all the densities considered in this work. 
There could, however,  exist a  spontaneous transition of a region of size $R_c$ 
to the crystallized state via tunneling. The value of $R_c$, and the rate of this 
transition, will be determined, among other things,  by 
tension of the interface between the condensate and crystal state, 
which is hard  to evaluate. However, for estimates the following qualitative
arguments should suffice: the height of the barrier  for each particle is
$(Z e^2 J_0/m_H)^{1/2}=m_\gamma$, while  the number of bosons in the $R_c$ 
region $\sim R_c^3 J_0/Z$. Hence, the transition rate should scale as 
${\rm exp}(-m_\gamma J_0  R^4_c/Z)$. Since we expect that  $R_c> 1/m_\gamma$, 
the rate is strongly suppressed for the parameters at hand.

\subsection*{Effective field theory description}

We use a low-energy effective  field theory description 
to study  the charged condensate. Even though realistic temperatures in the system may be  
well above zero, we focus on the  zero-temperature limit.
The relevance of this limit  is  justified  {\it a posteriori} and goes 
as follows:  the spin-0 nuclei undergo the condensation to the zero-momentum state;  
their phonons  cannot be excited  since their gap, $m_\gamma$,  is 
greater  than $T_c$.  On the other hand, gap-less 
near-the-Fermi-surface quasielectrons will be excited.
Therefore, all the thermal fluctuations 
will end up being stored in the  fermionic quasiparticles. 
For the latter, however, the finite temperature effects aren't 
significant since their Fermi energy is so much higher,  
$T/J_0^{1/3}\ll 10^{-2}$.  We note that the finite  temperature effects, 
in a general setup with  condensed bosons, were calculated 
in Refs. \cite {Dolgov,Dolgov2}.

We begin  at scales that are well below the heavy mass scale
$m_H$, but  somewhat above the scale set by  
${\rm max}[\mu_f,  m_e]$, where $\mu_f$  and $m_e$ are the electron chemical potential 
and mass respectively. Hence the electrons are described  
by their Dirac Lagrangian, while for the  description of 
the nuclei we will use a charged scalar {\it order parameter} $\Phi(x)$.
As it was shown in \cite {GGRReff}, in a non-relativistic  
approximation for the nuclei,  the  effective  Lagrangian proposed by  
Greiter, Wilczek and Witten  (GWW) \cite {GWW} in 
a context of superconductivity, is also applicable here, 
given that  an appropriate reinterpretation of its variables 
and  parameters is made.

The construction of the  GWW  Lagrangian  is based on the following  
fundamental principles: it is  consistent  with the translational, 
rotational, Galilean and the global $U(1)$ symmetries, 
preserves the algebraic relation  between the charged current density 
and momentum density, gives the Schr\"odinger equation for the 
order parameter in  the lowest order, and is gauge invariant \cite {GWW}.  
Combined  with  the electron dynamics the GWW  Lagrangian reads 
(we omit for simplicity the Maxwell term):
\beq
{\cal L}_{eff} = {\cal P} \left (  
 {i\over 2} ( \Phi^*  D_0 \Phi -  (D_0 \Phi)^* \Phi)-
{| D_j  \Phi|^2  \over 2m_H} \right )\,+
{\bar \psi}(i\gamma^\mu D^f_\mu -m_f)\psi,
\label{Leff}
\eeq
where we use the standard notations for covariant derivatives with the 
appropriate charge assignments: $D_0 \equiv (\partial_0  - iZe A_0)$, 
$ D_j \equiv ( \partial_j - i Ze A_j) $,
$D^f_\mu = \partial_\mu +ie A_\mu $, 
while  ${\cal P}(x)$ stands for a general polynomial function of its argument.
The coefficients of this polynomial, ${\cal P}(x) = \sum^{\infty}_{n=0} C_n x^n $,
are dimensionful parameters that are inversely proportional to powers of 
a short-distance  cutoff of the effective  field theory\footnote{In general one should also 
add to the Lagrangian terms $\mu_{NR} \Phi^*\Phi$,  
$\lambda (\Phi^*\Phi)^2/m_H^2$, $\lambda_1 (\Phi^*\Phi){\bar \psi}\psi/(m_H J_0^{1/3})$,   
and other  higher dimensional operators that are consistent 
with all the symmetries and conditions that lead to (\ref {Leff}) 
(the Yukawa term is not). Here $\mu_{NR}$ denotes a non-relativistic chemical potential for 
the scalars.  These terms are not important for 
the low-temperature spectrum of small perturbations  we're interested in, 
as long as  $\lambda,\lambda_1 \lsim 1$ and  $J_0\ll m^3_H$.  
However, near  the phase transition point it is the sign of $\mu_{NR}$ that 
would  distinguish between the broken and symmetric phases,  so 
these terms should be included for the discussion of the symmetry 
restoration.  We also note that the scalar part of  (\ref {Leff}) is somewhat 
similar to the Ginzburg-Landau (GL) Lagrangian for superconductivity. However, 
there are significant differences between them, one such difference being that  the 
coherence length  in the GL theory is many orders of magnitude greater than the  
average interelectron separation, while in the present case, the ``size of the 
scalar''  $\Phi$ is smaller that the average interparticle 
distance.}. 

Once the basic Lagrangian is fixed,  we  introduce  the 
electron chemical potential term $\mu_{f} \psi^+ \psi\,$, 
where  $\mu_{f}=\epsilon_F=[(3 \pi^2 J_0)^{2/3}+m_f^2]^{1/2}$. 
This is the only term that  at the tree level sets a frame in which the electron total 
momentum is zero. 

There exists  a homogeneous solution of  the equations of motion 
that follow from the effective Lagrangian (\ref {Leff}) \cite {GGRRwd}:
\beq
Z|\Phi|^2 = J_0\,,~~~ A_\mu=0,~~~~{\cal P}^{\prime}(0)=1\,.
\label{sol}
\eeq 
(We use  the unitary gauge $\Phi = |\Phi|$).
The condition ${\cal P}^{\prime}(0)=1$ is satisfied by any 
polynomial functions  ${\cal P} (x)$ for which the first coefficient is 
normalized to unity
\beq
{\cal P} (x) = x + C_2 x^2+...\,.
\label{px}
\eeq
The above solution describes a neutral system of 
negatively  charged electrons  of charge density 
$- eJ_0$,  and positively  charged scalar  
condensate  of charge density $Ze\Phi^+\Phi=eJ_0$ 
\cite {GGRReff,GGDP}.

Calculation of the spectrum of small perturbations 
is straightforward. The Lagrangian density for the fluctuations 
in the quadratic approximation reads \cite {GGRR1}
\beq
{\cal L}_{2}= -{1\over 4} F_{\mu\nu}^2 +
{1\over 2} m_0^2 A_0^2 - {1\over 2} m_\gamma^2 A_j^2+
{1\over 2} \,A_0 
{(2m_Hm_\gamma)^2\over - \Delta}A_0\,,
\label{Ltau}
\eeq
where $\Delta$ denotes the Laplacian, and the 
last term emerged due to mixing of $A_0$ with the 
fluctuation  of the $|\Phi|$,  which we integrated out. As before,
\beq
m_\gamma^2 \equiv  {Ze^2 J_0\over m_H}\,,
\label{photonmass}
\eeq
and $m_0^2=m_\gamma^2+ C_2e^2 J^{2}_0$.  At this stage we retained  
the fermionic fluctuations only in the Thomas-Fermi approximation \cite {GGRRwd}; 
an important refinement of this approximation, discussed in  \cite {GGRReff}, 
will be included below.

That there are no pathologies in (\ref {Ltau}), such as ghost and/or tachyons,  
can be seen by calculating the Hamiltonian density:
\beq
{\cal H}= {\pi_j^2  \over 2} 
+ {F^2_{ij}\over 4}  + 
{1\over 2} 
(\partial_j \pi_j) \left ( m_0^2 + {4M^4 \over -\Delta} \right )^{-1}
(\partial_j \pi_j)+ {1\over 2}
m_\gamma^2A_j^2\,.
\label{ham}
\eeq
Here, $M^2\equiv m_Hm_\gamma$ and  $\pi_j \equiv -F_{0j}$. 
The Hamiltonian is positive semi-definite. Moreover, the spectrum has a mass gap
determined by $m_\gamma$ (\ref {photonmass}). There are two transverse polarizations 
of a massive photon, as well as the longitudinal mode, the phonon,  
with the same mass $m_\gamma$ \cite {GGRR1}.

The massive bosonic collective excitations give rise to exponentially suppressed  
contributions to the value of specific heat of the charged condensate 
since typically $m_\gamma\gg  T_c$. The suppression scales as  ${\rm exp} (-m_\gamma /T)$,  
where $T\lsim T_c$. This is in contrast with the crystal, 
where the dominant contribution to the specific heat comes from a gap-less phonon, 
and scales with temperature as $T^3$.

As to the electrons, their behavior is similar in  both crystal 
and condensate  cases. At temperatures of interest  they form a 
degenerate Fermi gas  with gap-less excitations near the Fermi surface. 
Their contribution to the specific heat scales
linearly with temperature. In the case of crystallized substance   
this is sub-dominant to the specific heat due to the crystal phonon. 
For the charged condensate, however, the (quasi)electron fluctuations are 
the dominant contributors to  the specific heat.

\vspace{0.3cm}

To study the effects of collective bosonic and fermionic modes, as an interesting 
example, we  look at a potential  
between  two impurity nuclei (say hydrogen, or helium-3) 
of charge  $Q_1$ and $Q_2$. The calculation of the propagator that  
involves the light collective modes (for relativistic fermions) 
gives the following result \cite {GGRReff}:
\beq
V_{stat}= \alpha_{\rm em}  {Q_1Q_2} \left (  { e^{-Mr}\over \,r}
{\rm cos} (Mr)\, + {4 \alpha_{\rm em} 
\over \pi} { k_F^5{\rm sin}(2k_Fr)\over M^8r^4}\right )\,. 
\label{potential}
\eeq
The  first, exponentially suppressed term modulated by a periodic function, 
is due to cancellation between the screened Coulomb potential and 
that of a phonon \cite {GGRReff}.  The fact  of such a cancellation, and that it could give rise
to the oscillatory behavior of the exponentially screened potential was 
pointed out before in Ref. \cite {Lee} in the context of superconductivity\footnote{I'd like to 
thank Ki-Myeong Lee who  recently brought the  paper \cite {Lee} to my attention.}.

Most important, however, is the second term in (\ref {potential}) that 
has a long-range \cite {GGRReff}.  It dominates 
over the exponentially suppressed  term in (\ref {potential}) 
for scales of physical interest, and exhibits the  power-like behavior 
modulated by a periodic function. 

The potential  (\ref {potential}) is a generalization of the Friedel potential 
to  the case when in addition to the fermionic excitations  there are also  
collective modes due to   the  charged condensate.
The long-range oscillating term  in (\ref {potential})
is also a result of a subtraction between the conventional Friedel term and 
the long-range oscillating term due to a phonon. As a 
result, its magnitude is suppressed   compared  to what it would have 
been in a theory without the condensed charged bosons \cite {GGRReff}
(see, \cite {Fetter} for the discussion of the conventional 
Friedel potential,  and Ref. \cite {Dolgov2} for its recent  
detailed study in the presence of the charged condensate at finite 
temperature.)\footnote{Note 
that for  spin-dependent interactions the same effects of the charged condensate 
would give a generalization of the  Ruderman-Kittel-Kasuya-Yosida 
(RKKY) potential  \cite {RKKY}.}

The potential (\ref {potential}) 
is not sign-definite.  In particular, it can  give rise to 
attraction between like  charges;  this attraction is due to  
collective excitations of both  fermionic and bosonic degrees of freedom.  
This represents a generalization of the Kohn-Luttinger \cite {Kohn} 
effect to the case where on top of  the fermionic excitations  
the  collective modes of the  charged condensate 
are also contributing\footnote{In the charged condensate  
Cooper pairs of electrons can also be formed,  however, the 
corresponding transition temperature, and 
the magnitude of the gap, are  suppressed by a factor 
${\rm exp} (-1/e_{eff}^2)$, where $e_{eff}^2$  is proportional to 
the value of the inter-electron potential that contains both screened 
Coulomb and phonon exchange. The fact that this potential has attractive 
domain, but is  very  small,   is suggested by  
the static potential found in   
\cite {GGRReff} (see also eq. (\ref {potential}) above); 
the latter is suppressed by a power of 
a large scale $M$.  Furthermore, taking into account the 
frequency dependence  of the propagator in the Eliashberg equation does 
not seem to change qualitatively the conclusion on a strong  
suppression of the Green's function and pairing temperature.  

Hence, even though the bosonic  sector (condensed nuclei) is superconducting 
at  reasonably high  temperatures $\lsim 10^{6}~K$, 
interactions with gap-less fermions could dissipate the superconducting 
currents. Only at extremely low temperatures, exponentially close 
to the absolute zero, the electrons could also form a  gap leading 
to superconductivity of the whole system. In the present work we 
consider temperatures at which electrons are not  condensed into  
Cooper pairs, and ignore the finite temperature effects.\label{footnote}}.

\subsection*{Applications to White Dwarfs}

The above described system of electrons and nuclei  constitutes cores of 
white dwarf stars. Up to a factor of a few, these are 
roughly Earth's size  solar-mass objects; their  
mass  density  may range over  $\sim (10^6-10^{10})~g/cm^3$, 
most of them being near the lower edge of this interval.
Since  the dwarf stars exhausted  thermonuclear fuel in their cores 
already, they evolve by cooling \cite {Mestel};   
the ones that we consider in this work  cool from $\sim 10^7~K$ 
down to lower temperatures.

As a typical dwarf star cools down,
the Coulomb interaction  energy in a classical plasma of  
charged nuclei  will significantly  exceed their classical thermal energy, 
and the nuclei, in order to minimize energy,  
would organize themselves into a crystal lattice 
\cite {MestelRuderman}. In most of these  cases  
quantum effects of the  nuclei should be negligible;  
for instance, the Debye temperature should  be less than the 
temperature at which crystallization takes place, and  
the de Broglie wavelengths of the nuclei should be much 
smaller than  the average internuclear separations. 
This indeed is the case in majority of white dwarf stars,  
the cores of which  are  composed of carbon and/or oxygen nuclei 
and span the interval of mass densities around $\sim ( 10^6-10^8) ~g/cm^3$.

However, there exists a class of dwarf stars in which the nuclei
enter the quantum regime before the classical 
crystallization process sets in \cite {Ashcroft,Chabrier}.
Among these, furthermore, there is a relatively small 
subclass of the dwarf stars for which the temperature $T_c$,   
is  higher than the would-be crystallization temperature $T_{cryst}$ 
\cite {GGDP}. In such dwarf cores the charged condensation should  
be expected to take place.

White dwarfs composed of helium constitute a smaller sub-class of 
dwarf stars (see, \cite {Eis,24} are references therein);  they 
exhibit  best conditions for the charged  condensation. 
Most of helium dwarfs are believed to be formed in binary systems, where 
the removal of the envelope  off  the dwarf 
progenitor red giant by its binary companion happened 
before helium ignition, producing a remnant that evolves to 
a white dwarf with a helium core. Helium dwarf masses range from $\sim 0.5~M_\odot$ 
down to as low as $(0.18-0.19)~M_\odot$, while their envelopes are 
mainly composed of hydrogen. 

Using the approach of \cite{Shapiro}, and following \cite {GGDP} 
we will consider an over-simplified model of 
a reference helium star of mass $M=0.5~M_\odot$ with the 
atmospheric mass fractions of the hydrogen, and heavy elements 
(metallicity) respectively equal to
\beq
 X\simeq 0.99, \quad  \quad Z_m \simeq (0.0002-0.002)~.
\label{eq01}
\eeq
The lower value of the metallicity $ Z_m \simeq 0.0002$ 
is appropriate for the recently discovered 24 He WDs in 
NGC 6397 \cite {24}, but for completeness, we consider a 
wider range for this parameter.

It is straightforward to find the following expression for the 
cooling time of a star in the classical regime \cite {Shapiro}
\beq
t_{He}=\frac{k_B}{CAm_u}\left [ \frac{3}{5}(T_f^{-\frac{5}{2}}-
T_0^{-\frac{5}{2}})+Z\frac{\pi^2}{3}\frac{k_B}{E_F}
(T_f^{-\frac{3}{2}}-T_0^{-\frac{3}{2}})\right ],
\label{eq03}
\eeq
where $T_f$ and $T_0$ denote the final and initial core 
temperatures. The first term in the bracket on the right hand side 
corresponds  to cooling due to classical gas of the ions and the 
second term corresponds to the contribution coming from the 
Fermi sea. The latter  is sub-dominant in the range of final 
temperatures we are  interested in (the factor Z in front of this 
term is due to  $Z$ electrons per ion). Since  
$T_ f\ll T_0$, the age of a dwarf star typically  doesn't depend on 
the initial temperature.  Neglecting the fermion contribution, we find time 
that is needed to cool down to critical temperature $T_f=T_c$
\beq
t_{He}=\frac{3}{5}\frac{k_BT_c M}{Am_uL(T_c)}\simeq (0.76 - 7.6)~\text{Gyr}\,.
\label{ages}
\eeq
Where an order of magnitude interval in (\ref {ages}) 
is due to the interval in  the envelope  metallicity 
composition given in (\ref {eq01}).  We also find  
the corresponding luminosities
\beq
L(T_c) \simeq(10^8~erg/s)\frac{M}{M_\odot}
\left ( \frac{T_c}{\text{K}}\right )^{{7/2} }\simeq 1.5\cdot 
(10^{-4}-10^{-5}) L_{\odot}\,,
\eeq
which are in the range of observable luminosities ($L_{\odot}\simeq
3.84\cdot 10^{33} ~erg/s$).

After the condensation, specific heat of the system dramatically drops 
as the collective excitations of the condensed nuclei become 
massive and  ``get extinct''.  A contribution from the Fermi sea, which is 
strongly suppressed by the value of Fermi energy, becomes the dominant one.   
The phase transition itself would take some time to 
complete, and the drop-off in specific heat will not be instantaneous.

In the zeroth approximation, we can regard the transition to be  
very fast, and retain only the fermion 
contribution to specific heat below $T_c$. Then, the expression for the age of 
the star for $T_f<T_c$, reads as follows
\beq
t_{He}'=\frac{k_B}{CAm_u}\left [ \frac{3}{5}(T_c^{-\frac{5}{2}}-
T_0^{-\frac{5}{2}})+Z\frac{\pi^2}{3}\frac{k_B}{E_F}
(T_f^{-\frac{3}{2}}-T_0^{-\frac{3}{2}})\right ].
\label{eq02}
\eeq
Notice the difference of (\ref {eq02}) from  (\ref {eq03}) -- 
in the former $T_f <T_c$ and it is $T_f$ that enters as final temperature 
in the  fermionic part, while  $T_c$ should be taken as the 
final temperature in the bosonic part. 

From the ratio of ages, $\eta= {t_{He}/ t_{He}'}$, 
for two identical helium dwarf stars, with and without the 
interior condensation, we deduce  that the charged condensation substantially 
increases the rate of  cooling-- the age could be twenty times less 
than it would have been without the condensation phase \cite {GGDP}.

The condensation of the core would induce significant deviations from the 
classical curve for helium white dwarfs. What is independent of the uncertainties 
involved in these discussions, is the 
fact that the luminosity function (LF) will experience a significant drop-off after the 
charged condensation phase transition is complete. This is due to 
the ``extinction''  of the bosonic quasiparticles below the phase 
transition point. In fact, the LF   will drop by a factor of 
$\sim 200$. This may be relevant for an explanation of the observed termination of 
a sequence of the 24 He WD's found in \cite {24}. See Ref. \cite {GGDP} for more details.

Finally, the magnetic properties of the charged condensate, which are similar to those 
of type II superconductor, and in particular admit  the presence of Abrikosov's  
vortices, were studied in Ref. \cite {GGRRmag}. As was shown there, only very strong 
magnetic fields,  $\gsim 10^7~Gauss$,  will be able to penetrate the dwarf cores in the 
vortices,  while  weaker fields will be entirely expelled from it.

\vspace{0.2in}

\begin{center}
{\bf   Acknowledgments}
\end{center}

The above-reported  results constitute a part of the work done 
in collaboration with Rachel A. Rosen and David Pirtskhalava 
\cite {GGRR1}-\cite{GGDP}, \cite {GGRRmag}.
I'd like to thank Paul Chaikin, Daniel Eisenstein,  Leonid Glazman, Andrei Gruzinov, 
Stefan Hofmann, Andrew MacFadyen, Juan Maldacena, Aditi Mitra, Slava Mukhanov, 
Hector Rubinstein, Malvin Ruderman and Arkady Vainshtein for 
useful discussions and  correspondence on these topics.
The work  was supported by  the NSF  grant PHY-0758032.

\end {document}